\documentclass[pra,showpacs,twocolumn,epsfig]{revtex4}
\usepackage{amsfonts}
\usepackage{amsmath}
\usepackage{amssymb}
\usepackage{graphicx}

\begin{document}

\title{Storage of spin squeezing in a two-component Bose-Einstein condensate}
\author{Guang-Ri Jin \cite{email} and Sang Wook Kim \cite{email1}}
\affiliation{Department of Physics Education and Department of
Physics, Pusan National University, Busan 609-735, Korea}
\date{\today }

\begin{abstract}
Efficient control of spin squeezing in a two-component Bose-Einstein
Condensate is studied by rapidly turning-off the external field at a
time that maximal spin squeezing appears. We show that strong
reduction of spin fluctuation can be maintained in a nearly fixed
direction for a long time. We explain the underlying physics
unambiguously, and present analytical expressions of the
maximal-squeezing time.
\end{abstract}

\maketitle

Spin squeezing has attracted much attention for decades not only
because of fundamental physical interests
\cite{Kitagawa,Wineland,Kuzmich,Hald,Geremia}, but also for its
possible application in atomic clocks for reducing quantum noise
\cite{Wineland} and quantum information
\cite{Sorensen,You,Wang,Lewenstein,Yi}. Formally, the spin
squeezing is quantified via a parameter $\xi =(\Delta
\hat{J}_{{\bf n}})_{\min}/\sqrt{j/2}$, where $(\Delta
\hat{J}_{{\bf n}})_{\min}$ represents the smallest variance of a
spin component $\hat{J}_{\bf n}=\hat{J}\cdot{\bf n}$ normal to the
mean spin $\langle \hat{J}\rangle$. For the coherent spin state
(CSS), the variance $(\Delta \hat{J}_{{\bf n}})_{\min
}=\sqrt{j/2}$ (i.e., $\xi=1$). A state is called spin squeezed
state (SSS) if its variance of the spin component is smaller than
that of the CSS, i.e. $\xi<1$.

Kitagawa and Uea have investigated the spin squeezing generated by
the so-called one-axis twisting (OAT) model with Hamiltonian:
$\hat{H}_{\text{OAT}}=2\kappa\hat{J}_z^2$ \cite{Kitagawa}.
Possible realization of the OAT-type spin squeezing in a
two-component Bose-Einstein Condensate (TBEC) has been proposed in
Refs. \cite{Sorensen,Molmer}. S\o rensen et al. also argued that
macroscopic quantum entanglement can be characterized by using the
OAT-type spin squeezing in the TBEC \cite{Sorensen}. Besides the
TBEC, most recently, Takeuchi et al. considered another
realization of the OAT-type spin squeezing by using the
interactions between atoms and off-resonant light (paramagnetic
Faraday rotation) \cite{Takeuchi}. To coherently control spin
squeezing, Law et al. introduced additional Josephson-like (or
Raman) coupling to the OAT model: $\hat{H}_R=
2\kappa\hat{J}_z^2+\Omega_R\hat{J}_x$ \cite{Law}. Such a model has
also been used to prepare arbitrary Dicke states \cite{Bigelow}.

In addition to the generation of the SSS itself, it is desirable
to maintain not only the squeezing but also its direction for a
long time \cite{You}. Jaksch et al. have shown that the OAT-type
SSS can be stored for arbitrarily long time by removing the
self-interaction \cite{Jaksch}. However, it might not be easy to
handle in experiment since the precisely designed additional
pulses are crucially required. In this letter, we propose a simple
mechanism to obtain long-lasting spin squeezing in the TBEC. Our
scheme is quite easy to realize in experiment since it can be
achieved by turning-off the Josephson coupling once the TBEC
reaches its maximal spin squeezing.

We consider a two-component weakly-interacting BEC
\cite{Hall,Stenger} consisting of $N$ atoms in different atomic
hyperfine states $|a\rangle$ and $|b\rangle $ coupled by a
time-varying microwave field with Rabi frequency $\Omega_{rf}$.
Based on the two-mode approximation
\cite{Milburn,Smerzi,TMA1,TMA2,TMA3,Savage99}, the total
Hamiltonian can be described by ($\hbar =1$):
\begin{equation}
\hat{H}(t)=2\kappa \hat{J}_{z}^{2}+\Omega (t)\hat{J}_{x},
\label{hamiltonian}
\end{equation}%
where $\kappa =(\kappa _{aa}+\kappa _{bb}-2\kappa _{ab})/4$, and
$\kappa _{\alpha \beta }=g_{\alpha \beta}\int d^{3}{\bf
r}\left\vert \phi _{\alpha }({\bf r})\phi_{\beta }({\bf
r})\right\vert ^{2}$, with $g_{\alpha \beta}=4\pi a_{\alpha \beta
}/m$ $(\alpha ,\beta =a,b)$ being the $s$-wave scattering
strengthes between atoms. The condensate-mode functions
$\phi_{\alpha}$ normalized to unity satisfies a coupled
Gross-Pitaevskii equations \cite{Castin}. Here we focus on the
case that the external field is turned off rapidly at a certain
time $t_0$, so the time-dependent Josephson-like coupling can be
written as $\Omega (t)=\Omega_R\Theta(t_0-t)$, where
$\Omega_R=\Omega_{rf}\int d^{3}{\bf r}\phi_{a}^*({\bf
r})\phi_{b}({\bf r})$ and $\Theta (t)$ is the step function.

The state vector at arbitrary time $t$ can be expanded in terms of
eigenstates of $\hat{J}_{z}$: $|\psi (t)\rangle
=\sum_{m}c_{m}(t)\left\vert j,m\right\rangle$, where $-j\leq m\leq
j$ and $j=N/2$. The equations of motion for the amplitudes
$c_{m}(t)$ are obtained by solving time-dependent Schr\"{o}dinger
equation. We consider an initial CSS $|j,-j\rangle_{x}=e^{-i\pi
J_{y}/2}|j,-j\rangle$, then the initial amplitudes
$c_{m}(0)=\frac{(-1)^{j+m}}{2^{j}}\left(
\begin{array}{c}
2j \\
j+m
\end{array}%
\right)$. Since the initial CSS satisfies $c_{-m}(0)=c_{m}(0)$ for
even $N$, and $c_{-m}(0)=-c_{m}(0)$ for odd $N$, we can prove the
mean spin always along the $x$ direction. In addition, we will
consider only positive $\kappa$ case by assuming
$a_{aa}+a_{bb}>2a_{ab}$. However, our results keep valid in the
opposite case by using initial maximum weight state of $\hat{J}_x$,
i.e., $|j,j\rangle_x$.

\begin{figure}[htbp]
\begin{center}
\includegraphics[width=8.5cm]{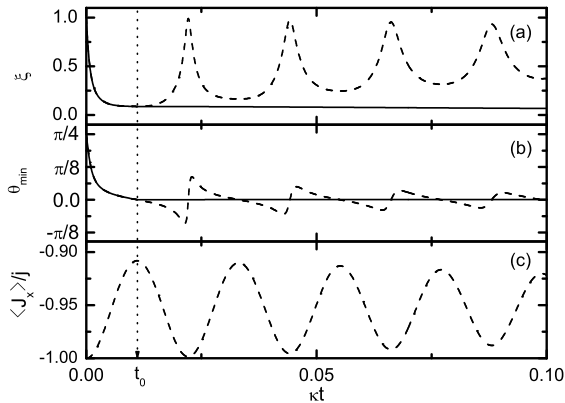}
\end{center}
\vskip -0.5cm \caption{Time evolution of (a) the squeezing
parameter, (b) the squeezing angle, and (c) the mean spin
$\langle\hat{J}_x\rangle$ for $N=10^3$ and $\Omega_R=10.8\kappa$.
Dashed lines: constant-coupling case; Solid lines: turning-off
external field at the time $\kappa t_0=1.1\times10^{-2}$, with its
position indicated by the vertical dotted line.}\label{fig1}
\end{figure}

Now let us briefly explain the basic principle of our scheme. The
initial CSS can be prepared by applying a short $\pi/2$ pulse to a
single-component BEC with all the atoms being in the internal state
$|a\rangle$ \cite{Sorensen,Savage99}. After that, the external
Josephson field is immediately switched on, so dynamical evolution
of the spin system is governed by the Hamiltonian
(\ref{hamiltonian}) with $\Omega(t)=\Omega_R$. If the coupling is
optimally chosen, the Josephson interaction results in an enhanced
spin squeezing compared with that of the OAT \cite{Law}. For
$N=10^3$, we find that the maximal squeezing $\xi_0=8.7076\times
10^{-2}$ can be obtained by choosing arbitrary $\Omega_R$ in a
region $10.777\leq\Omega_R/\kappa\leq10.818$. As shown by the dashed
lines of Fig.~\ref{fig1}, the squeezing $\xi$ and the mean spin
$\langle\hat{J}_x\rangle$ show collapsed oscillations
\cite{Agarwal}. At the time $t_0$, $\xi$ decreases to its local
minimum $\xi_0$ with $\theta_{\min}=0$, while
$\langle\hat{J}_x\rangle$ increases to its maximal value
$\langle\hat{J}_x\rangle_0$. The basic features of our scheme are
exhibited by the solid lines of Fig.~\ref{fig1}. We find that if we
turn off the Jesephson field at the time $t_0$, the maximal
squeezing $\xi_0$ can be stored in a fixed direction (i.e.,
$\theta_{\min}=0$) for a long time.

\vskip -0.5cm

\begin{figure}[htbp]
\begin{center}
\includegraphics[width=9.0cm]{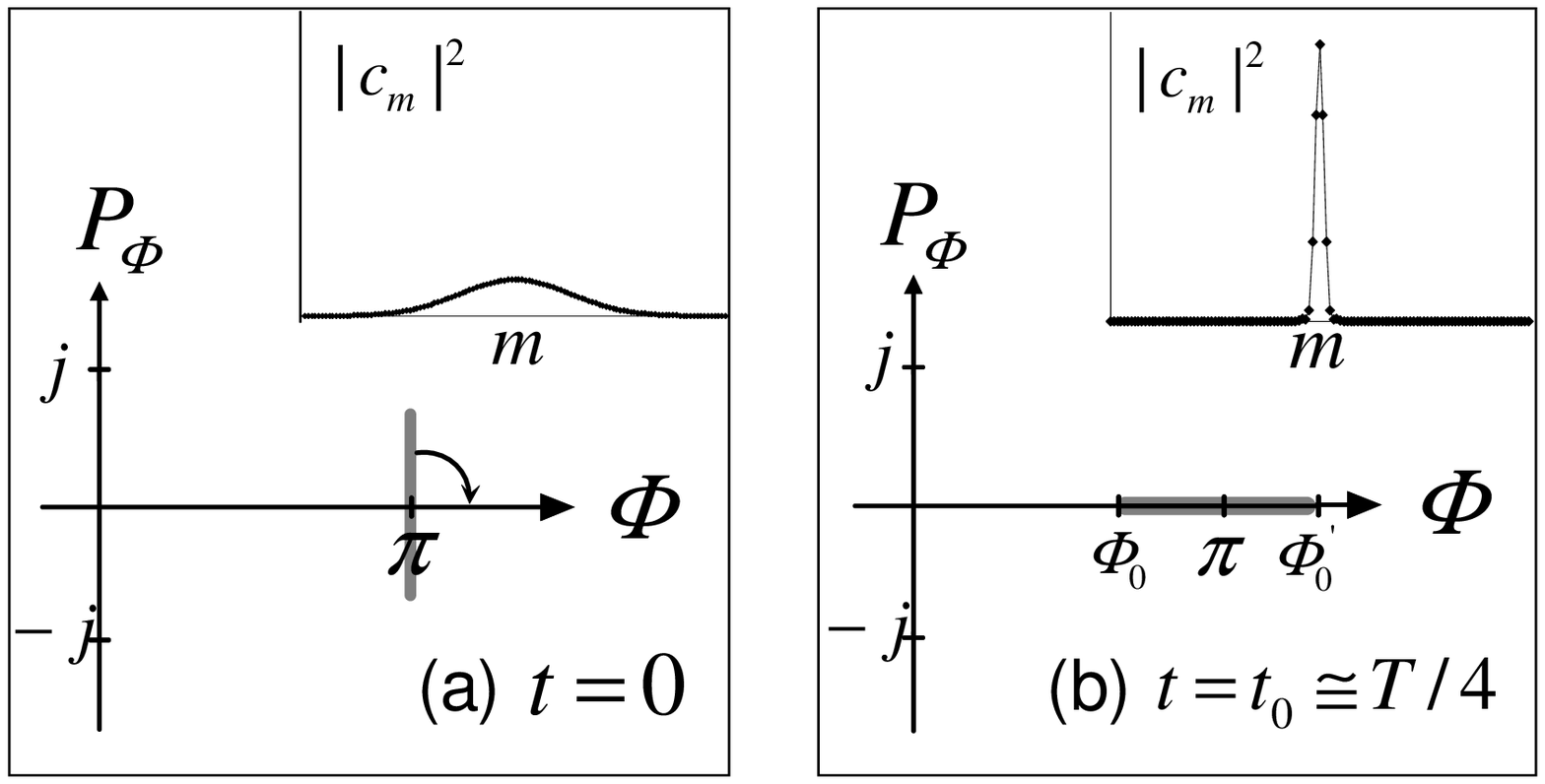}
\vskip -8.5cm \caption{Schematic picture of the probability
distribution in phase space ($\Phi$, $P_\Phi$) for: (a) the initial
CSS, (b) the SSS at $t\simeq T/4$. The insets: the corresponding
distribution $|c_m|^2$ as a function of $m$ (or $P_\Phi$) obtained
numerically. } \label{fig2}
\end{center}
\end{figure}

To understand our above observations, we investigate probability
distribution of the spin state, $|c_m|^2=|\langle
j,m|\psi(t)\rangle|^2$. As shown in the insets of Fig.~\ref{fig2},
we find that compared with the initial CSS, the maximally SSS at
$t_0$ has a very sharp probability distribution centered at the
lowest spin projection, i.e., $m=0$ (for even $N$) or $m=\pm1/2$
(for odd $N$). Such a sharp probability distribution of the SSS
can be explained qualitatively by considering the familiar phase
model \cite{phase model}. By replacing $\hat{J}_z \rightarrow
P_{\hat{\Phi}}=-i\partial_{\hat{\Phi}}$ and $\hat{J}_x \rightarrow
(N \cos \hat{\Phi})/2$, with $\hat{\Phi}$ being macroscopic phase
difference between two condensate components, one obtain
\begin{equation}
H_\Phi = -2\kappa\frac{\partial^2}{\partial\hat{\Phi}^2} +
\frac{\Omega_R N}{2}\cos{\hat{\Phi}},\label{phase_model}
\end{equation}
where we have taken $\Omega(t)=\Omega_R$ to simulate quantum
dynamics of the spin system before turning-off the field. The
phase model Hamiltonian allows us to regard the spin system as a
fictitious particle with effective mass $(4\kappa)^{-1}$ subject
to a pendulum potential. Moreover, the particle behaves as a
pendulum rotating with frequency
$\omega_{\text{eff}}=\sqrt{2\kappa\Omega_R N}$ in phase space
($\Phi$, $P_\Phi$), where mean phase difference
$\Phi=\langle\hat{\Phi}\rangle$. As shown in Fig.~\ref{fig2}(a)
and (b), starting from vertical distributed initial points
($\Phi=\pi$ and $P_\Phi=-j,-j+1,..., j$), one obtains the
distribution elongated horizontally at the time $t \simeq T/4$,
where $T=2\pi/\omega_{\text{eff}}$ is the period of the pendulum.
It is reasonable to assume that the projection of the distribution
along $\Phi$ will always be symmetrical to $\pi$, e.g.,
$\Phi'_0=2\pi-\Phi_0$ and $\langle\hat{J}_x\rangle_0=j \cos
(\Phi_0)$, as Fig.~\ref{fig2}(b).

\vskip 0.2cm

Based upon the phase model, we explain intuitively why the
obtained SSS has a sharp distribution with the lowest spin
projection being occupied predominantly. In addition, we find that
the sharp distribution accompanies with the maximal mean spin,
i.e., $\langle\hat{J}_x\rangle$: $-j\rightarrow
\langle\hat{J}_x\rangle_0$ as $\Phi: \pi\rightarrow\Phi_0$, thus
$(d\langle\hat{J}_x\rangle/dt)_{t_0}\equiv 0$. On the other hand,
from Heisenberg equation of $\hat{J}_x$ we know the relation
$d\langle\hat{J}_x\rangle/dt\sim \langle\hat{J}_{z}
\hat{J}_{y}+\hat{J}_{y}\hat{J}_{z}\rangle\sim
A\tan(2\theta_{\min})$ [for $A\neq0$, see the deduction of
Eq.~(6)], so we further obtain $\theta_{\min}=0$ at $t_0$, as
shown in Fig.~\ref{fig1}(b). More important, we obtain analytical
expression of the maximal-squeezing time
\begin{equation}
\kappa t_0 \simeq \kappa \frac{T}{4} =
\frac{\pi}{2}\sqrt{\frac{\kappa}{2\Omega_R N}},\label{t_M}
\end{equation}
which is valid for large $N$ ($\geq10^3$) and small coupling with
$\kappa<\Omega_R<<N\kappa$. By comparing with exact numerical
results, we find that Eq.~(\ref{t_M}) gives accurate prediction of
the maximal-squeezing time for $\Omega_R$ near or larger than the
optimal coupling. Law et al. have investigated the optimal
coupling as a function of $N$ based on a wide range of numerical
simulations \cite{Law}, from which we suppose that the optimal
coupling obeys power rule $\Omega_R/\kappa\sim N^{1/3}$ for the
large $N$. Such as $N=10^3$, the optimal coupling is about
$\Omega_R=10.8\kappa$ and Eq.~(\ref{t_M}) gives $\kappa
t_0=1.069\times10^{-2}$, consistent with numerical result
$1.104\times10^{-2}$. Note that Eq.~(\ref{t_M}) predicts the time
scale for the sharp distribution of the SSS and the maximal
$\langle\hat{J}_x\rangle$ (also $\theta_{\min}=0$), which,
however, does not necessarily correspond to the maximal squeezing
(see below).

\begin{figure}[htbp]
\begin{center}
\includegraphics[width=6cm, height=8.5cm, angle=270]{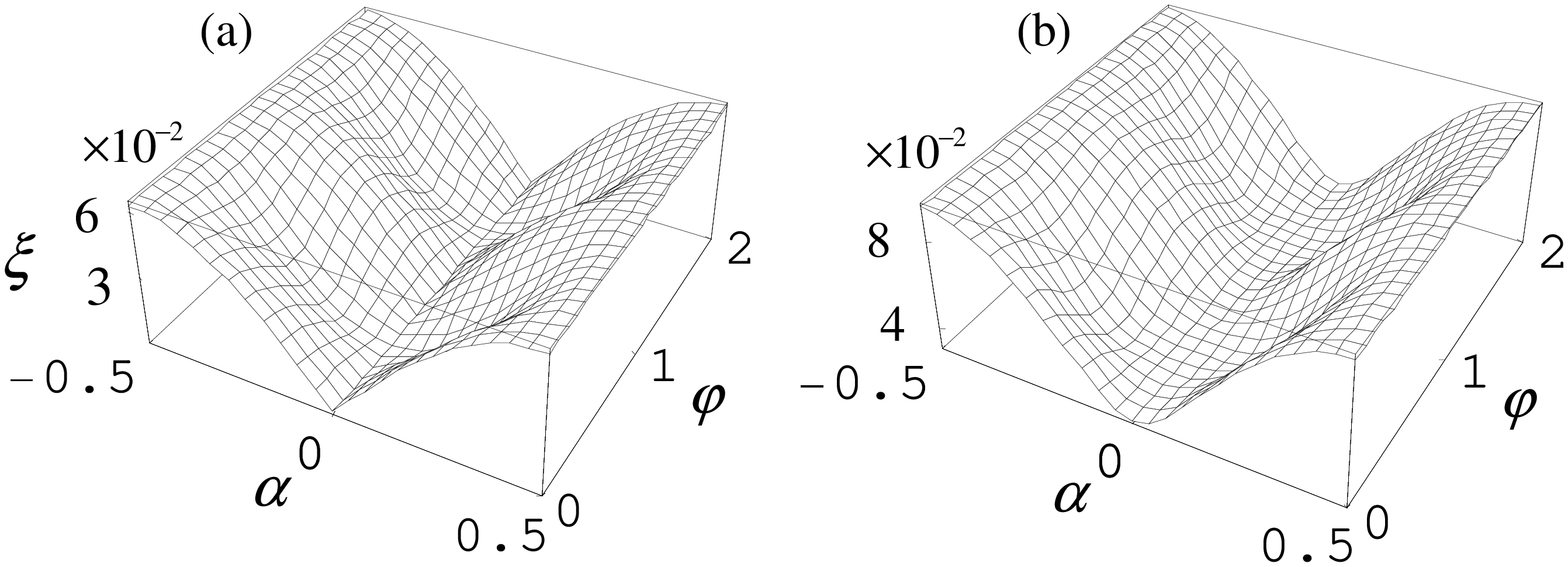}
\vskip -2.5cm \caption{The squeezing parameter $\xi$ as a function
of $\alpha$ and $\varphi$ (in units of $\pi$) for (a) the even
number case ($N=1000)$; (b) the odd number case ($N=1001$),
calculated by using Eqs. ~(\ref{evenN}) and (\ref{oddN}),
respectively.} \label{fig3}
\end{center}
\end{figure}

If the Josephson field is turned off at $t_0$, the spin system is
governed only by the self-interaction Hamiltonian $2\kappa J_z^2$,
so the distribution $\left|c_m(t)\right|^2$ is conserved while the
relative phases among the spin projections are subject to change.
When the SSS at the time $t_0$ exhibits a very sharp distribution
in $m$, as shown in the inset of Fig.~\ref{fig2}(b), the effect of
relative phases induced by the self-interaction gives negligible
influence to the squeezing. To show this, we suppose the SSS at
$t_0$ takes the form
\begin{equation}
|\psi(t_0)\rangle=\frac{e^{i\varphi}\sin\alpha}{\sqrt{2}}\left(\left\vert
j,1\right\rangle+\left\vert
j,-1\right\rangle\right)+\cos\alpha\left\vert
j,0\right\rangle,\label{evenN}
\end{equation}
for even $N$ case, or
\begin{eqnarray}
|\psi(t_0)\rangle&=&\frac{e^{i\varphi}\sin\alpha}{\sqrt{2}}\left(\left\vert
j,3/2\right\rangle -\left\vert j,-3/2\right\rangle\right)\notag\\
&&+\frac{\cos\alpha}{\sqrt{2}}\left(\left\vert j,1/2\right\rangle
-\left\vert j,-1/2\right\rangle\right),\label{oddN}
\end{eqnarray}
for the odd $N$, where $\varphi$ and $\alpha$ represent the
relative phase and the amplitude, respectively. Fig.~\ref{fig3}
shows the squeezing parameter $\xi$ as a function of $\alpha$ and
$\varphi$, where two distinct features are observed. Firstly,
$\xi$ is minimized as $\alpha\rightarrow 0$ \cite{Wineland}, which
implies that the maximal squeezing occurs as long as the SSS has a
very sharp distribution with a large amplitude of the lowest spin
projection or, equivalently, as the SSS approaches to the
ground-state of $2\kappa J_z^2$. Secondly, $\xi$ looks insensitive
to the relative phase $\varphi$ for the SSS with a sharp
distribution. Since the self-interaction only varies the relative
phase $\varphi$ with a fixed $\alpha$, Fig.~\ref{fig3} explains
qualitatively the storage of the maximal squeezing after $t_0$.

The validity of Eqs. (\ref{evenN}) and (\ref{oddN}) can be tested
by considering two exact solvable cases with $N=2$ and $N=3$.
Taking the optimal coupling $\Omega_R=\kappa$ for the two-atom
system and $\Omega_R=2\kappa$ for the three-atom one, we obtain
the maximally-squeezed states
$|\psi(t_{n})\rangle=i(-1)^{n+1}e^{-i\kappa t_{n}}\vert
j=1,m=0\rangle$ \cite{Wineland} and $|\psi (t_{n})\rangle
=\frac{i(-1)^n}{\sqrt{2}}e^{-3i\kappa t_{n}/2}\left(\vert
3/2,1/2\rangle -\vert 3/2,-1/2\rangle\right)$, respectively, where
$t_{n}=(2n+1)\pi /S$ for any integer $n$, and the level spacing
$S=2\sqrt{\Omega_R^2+\kappa^2}$ for $N=2$ and
$S=2\sqrt{\Omega_R^{2}+2\kappa \Omega_R+4\kappa^{2}}$ for $N=3$.
Obviously, the spin state $|\psi (t_{n})\rangle$ is the ground
state of the self-interaction Hamiltonian $2\kappa\hat{J}^2_z$,
which results in exactly constant $\xi$ with zero $\theta_{\min}$
by rapid turning-off the external field at the times $t_{n}$. For
large $N$, the SSS at the time $t_0$ no longer corresponds to the
ground state of $2\kappa\hat{J}^2_z$, but approaches to it
compared with the initial CSS. Consequently, almost constant $\xi$
can be achieved after turning-off the external field.

\begin{figure}[htbp]
\begin{center}
\includegraphics[width=16cm]{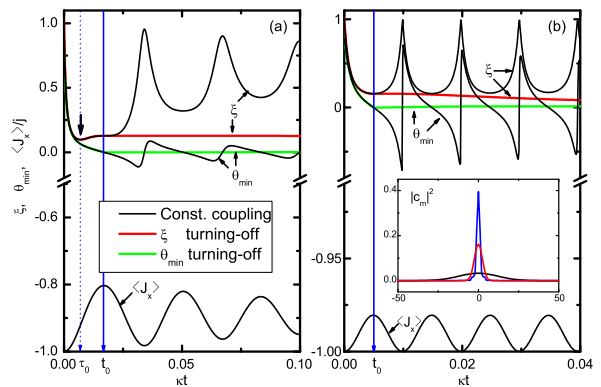}
\end{center}
\vskip -0.5cm \caption{(Color online) Time evolution of $\xi$,
$\theta_{\min}$ and the mean spin $\langle\hat{J}_x\rangle/j$ for
$N=10^3$, (a) $\Omega_R=5\kappa$, and (b) $\Omega_R=50\kappa$. Black
curves in (a) and (b) are the corresponding quantities for the
constant-coupling case. Red (green) curves represents the squeezing
$\xi$ ($\theta_{\min}$) after turning off the field at time $t_0$.
Inset: Probability distribution $|c_m|^2$ as a function of $m$ for
the SSS at the times $\tau_0$ (the black line), and $t_0$ with the
blue line for (a) and the red line for (b).}\label{fig4}
\end{figure}

The storage itself in our scheme does not depend on $\Omega_R$,
and the optimal coupling is chosen here to store the maximal
squeezing $\xi_0$. If $\Omega_R$ is smaller than the optimal
coupling, there exist two time scales: $t_0$ for the vanishing
$\theta_{\min}$, and $\tau_M$ for the maximal squeezing.
Eq.~(\ref{t_M}) still works well to give the time scale of
$\theta_{\min}=0$, but fails to predict that of the maximal
squeezing. As shown in Fig.~\ref{fig4}(a), for $\Omega_R=5\kappa$
real maximal squeezing occurs at $\kappa
\tau_0=6.915\times10^{-3}$. From Fig.~\ref{fig4}(a) we also find
that the maximal $\langle\hat{J}_x\rangle$ and $\theta_{\min}=0$
appears at the same time $\kappa t_0=1.687\times10^{-2}$, at which
the probability distribution of the SSS is sharp enough [see the
inset of Fig.~\ref{fig4}(b)]. As a result, a less squeezed
variance with $\theta_{\min}=0$ can be stored by turning-off the
external field at the time $t_0$ [see the red and the green lines
of Fig.~\ref{fig4}(a)].

\vskip -0.5cm

It is worth mentioning that the above qualitative explanation can
not be applied to the SSS with a broad probability distribution.
In fact, typical OAT-scheme \cite{Kitagawa} relies solely on the
evolution of relative phases induced by the self-interaction,
where the initial CSS shows a broad probability distribution. As
shown by the red line of Fig.~\ref{fig4}(b), for a large coupling
$\Omega_R=50\kappa$, the squeezing parameter decreases slightly
after turning-off the external coupling. This is because the SSS
at $t_0$ exhibits a relatively broad probability distribution [see
the inset of Fig.~\ref{fig4}(b)]. From the green line of
Fig.~\ref{fig4}(b), we find that the reduced variance is stored in
the fixed direction with $\theta_{\min}=0$. By comparing the
numerical and the analytical result of $t_0$ for
$\Omega_R=50\kappa$, we also find that the numerical result
$4.945\times 10^{-3} \kappa^{-1}$ agrees very well with $4.967
\times10^{-3} \kappa^{-1}$ estimated from Eq.~(\ref{t_M}).

Before closing, we prove that the mean spin always appears in the
$x$ direction, and explain why we study the spin squeezing in the
small-coupling regime. Note that the linear combinations of the
probability amplitudes $p^{(\pm)}_{m}(t)=c_{m}(t)\pm c_{-m}(t)$
obey two closed sets of first-order ordinary differential
equations. For even $N$, the fact that all $p^{(-)}_{m}(0)=0$
results in $p^{(-)}_{m}(t)=0$, namely $c_{-m}(t)=c_{m}(t)$. On the
other hands, for odd $N$ all $p^{(+)}_{m}(t)$ are zero, i.e.,
$c_{-m}(t)=-c_{m}(t)$. Since $c_{-m}(t)=\pm c_{m}(t)$, we obtain
simple expressions: $\langle \hat{J}_{y}\rangle =\langle
\hat{J}_{z}\rangle =0$, and $\langle \hat{J}_{x}\rangle \neq 0$,
i.e., the mean spin is along the $x$ direction. Consequently, the
spin component normal to the mean spin reads $\hat{J}_{{\bf
n}}=\hat{J}_{y}\sin \theta +\hat{J}_{z}\cos \theta$. By minimizing
the variance $(\Delta \hat{J}_{\mathbf{n}})^{2}$ with respect to
$\theta$, we obtain the squeezing angle
$\theta_{\min}=\frac{1}{2}\tan^{-1}(B/A)$ and $(\Delta
\hat{J}_{{\bf n}})_{\min }^{2}
=\frac{1}{2}C-\frac{1}{2}\sqrt{A^{2}+B^{2}}$, where $A=\langle
\hat{J}_{z}^{2}- \hat{J}_{y}^{2}\rangle$, $B=\langle \hat{J}_{z}
\hat{J}_{y}+\hat{J}_{y}\hat{J}_{z}\rangle$, and
$C=\langle\hat{J}_{z}^{2}+\hat{J}_{y}^{2}\rangle$. From Heisenberg
equations of motion of the spin $\hat{J}_{\alpha}$ for
$\alpha=x,y,z$, one can obtain formal solutions for the constant
coupling case: $C=j(j+1)-\langle\hat{J}_x^2\rangle$, $A=-C +
j(1-\Omega_R/\kappa)-(\Omega_R/\kappa) \langle\hat{J}_x\rangle$,
and $B = -(2\kappa)^{-1}d\langle\hat{J}_x\rangle/dt$. Note that
for $B=0$ and $A\neq0$, the spin squeezing takes place along $z$
axis (i.e. $\theta_{\min}=0$) with the corresponding squeezing
parameter
\begin{equation}
\xi^2_0=1-(\Omega_R/\kappa)\left[1+\langle\hat{J}_x\rangle_0/j\right],
\end{equation}
where $\langle\hat{J}_x\rangle_0$ is the maximum value of the mean
spin. For an extremely strong coupling ($\Omega_R>>\kappa N$),
$\langle\hat{J}_x\rangle_0\rightarrow-j$ and $\xi_0\rightarrow1$
so the squeezing becomes very weak. This is the reason why we
discuss the spin squeezing in the small-coupling regime
($\kappa<\Omega_R<<N\kappa$).

Finally, we estimate several important parameters for experimental
realization. Following Ref. \cite{Sorensen}, we consider $^{23}$Na
atoms in the hyperfine states $|F=1, M_F=\pm1\rangle$ trapped in a
spherically symmetric potential $V_a=V_b=m\omega^2r^2/2$. The
self-interaction strength can be solved by applying the
Thomas-Fermi approximation, yielding
\begin{equation}
\kappa \simeq \frac{15^{2/5}\hbar\omega}{14}
\frac{a_{\text{eff}}}{a_{\text{ho}}}
\left(\frac{a_{\text{ho}}}{Na_{aa}}\right)^{3/5},
\end{equation}
where $a_{\text{ho}}=\sqrt{\hbar/m\omega}$ being the harmonic
oscillator length, and $a_{\text{eff}}=a_{aa}+a_{bb}-2a_{ab}$
effective scattering length. For $^{23}$Na atoms, we take
$a_{aa}=a_{bb}=2a_{ab}$ \cite{Sorensen} and
$a_{\text{eff}}=a_{aa}=2.75\text{nm}$ \cite{Stenger}, then the
self-interaction strength
$\kappa\simeq3.2448\times10^{-4}\hbar\omega$. For the case
$N=10^3$ and $\Omega_R=10.8\kappa$, we have obtained
$t_0=1.1041\times10^{-2}/(\hbar^{-1}\kappa)=34.03\omega^{-1}$,
which corresponds to the maximal-squeezing time about $10.83$ ms
for $\omega=2\pi\times500$Hz.

In summary, we have investigated coherent control of spin
squeezing by turning-off Josephson field at the maximal-squeezing
time. We show that by applying the optimal coupling then turning
off later, the maximal squeezing can be stored in the $z$ axis for
a long time, which can be explained in terms of the probability
distribution of the SSS. For a sharp distribution with a large
amplitude of the lowest spin projection, the effect of the
self-interaction gives small contribution to the squeezing and its
direction. We find the analytic expression of the
maximal-squeezing time by considering the phase model. Our scheme
for the storage of spin squeezing is quite robust for a wide range
of Josephson coupling.

We thank Profs. C. K. Kim, K. Nahm, C. P.
Sun, W. M. Liu, S. Yi, and X. Wang for helpful discussions. This
study was financially supported by Pusan National University in
program Post-Doc 2006 and Korea Research Foundation Grant funded by
the Korean Government (MOEHRD, Basic Research Promotion Fund)
(KRF-2006-312-C00543).


\end{document}